\documentclass[prd,amsmath,amssymb,superscriptaddress,nofootinbib]{revtex4-2}
\usepackage{amsfonts}
\usepackage{multirow}
\usepackage{makecell}
\usepackage{mathrsfs}
\usepackage{graphicx}
\usepackage{amsmath}
\usepackage{amssymb}
\usepackage{bm}
\usepackage{bbm}
\usepackage{color}
\usepackage{array}

\makeatletter

\begin{document}

\title{Exclusive Decays of the Fully Heavy Tetraquarks into Light Mesons}
\author{Feng Feng}
\email{f.feng@outlook.com}
\affiliation{China University of Mining and Technology, Beijing 100083, China\vspace{0.2cm}}
\author{Ming-Ming Liu}
\email{juliuslmm@outlook.com}
\affiliation{China University of Mining and Technology, Beijing 100083, China\vspace{0.2cm}}
\date{\today}

\begin{abstract}
In this work, we investigate the exclusive decays of the fully heavy tetraquark states $T_{4c,b}$ into light mesons, specifically $\pi$ and $K$, using the framework of Non-Relativistic QCD (NRQCD) and collinear QCD factorization for hard exclusive processes. We estimate the decay widths to be $10^{-9}$ GeV and $10^{-14}$ GeV for the decays $T_{4c} \to \pi^+\pi^-$ and $T_{4b} \to \pi^+\pi^-$ (and similarly for $K^+K^-$), respectively. The branching ratio for $T_{4c} \to \pi^+\pi^-(K^+K^-)$ is on the order of $10^{-8}$, making it currently unobservable in existing experiments.
The factorization of $T_{4c,b}$ into light hadrons shares similarities with the decay $J/\psi \to p\bar{p}$. However, unlike the latter process, the decay $T_{4c,b} \to \pi^+\pi^-(K^+K^-)$ exhibits unique features that arise only in processes involving multiple incoming or outgoing particles. One such feature is the necessity of maintaining the $i\varepsilon$-prescription for the denominators or propagators due to the divergences in the kinematic region of interest.
Employing the sector decomposition method, along with the aid of the Cheng-Wu theorem and the {\tt QHull} program, we present a systematic approach to handle the convolutions and phase-space integrations. This method can also be extended to similar processes, such as $T_{4c,b} \to p\bar{p}$, as well as to phase-space integrations where each denominator is a linear combination of integration variables.
\end{abstract}

\maketitle

\section{Introduction}  

In recent years, experimental searches for all-charmed tetraquarks have progressed significantly.
In 2020, the LHCb Collaboration reported a narrow structure, denoted as X(6900), with the significance larger than $5\sigma$~\cite{LHCb:2020bwg}, together with a broad enhancement in the di-$J/\psi$ invariant-mass spectrum around the double $J/\psi$ threshold region.
Subsequently, the ATLAS and CMS Collaborations confirmed the existence of X(6900) 
and observed additional broad structures in the di-$J/\psi$ and
$J/\psi\psi(2S)$ channels~\cite{ATLAS:2023bft,CMS:2023owd}. 
More recently, both the ATLAS and CMS Collaborations measured the ratio of partial decay widths
between $J/\psi\psi(2S)$ and di-$J/\psi$ channels, determining the quantum numbers of the X(6900) to be
${\rm J}^{\rm PC} = 2^{\rm ++}$~\cite{ATLAS:2025nsd,CMS:2025fpt}.
In the bottom quark sector, the CMS Collaboration reported an observation of a $\Upsilon$ pair~\cite{CMS:2016liw}; however, this structure was not yet confirmed by subsequent analyses from CMS~\cite{CMS:2020qwa}. Additionally, the LHCb Collaboration investigated the $\Upsilon\mu^+\mu^-$ invariant mass spectrum but did not observe any reliable signals~\cite{LHCb:2018uwm}.

Although the search for exotic states presents numerous challenges, the potential existence of exotic hadrons—especially those composed of four or five quarks—has been extensively studied in the literature (for recent reviews, see, e.g., Refs.~\cite{Liu:2019zoy,Chen:2022asf}). The pioneering ideas regarding fully charmed tetraquark configurations near the $J/\psi$-pair threshold can be traced back to the 1970s and 1980s~\cite{Iwasaki:1975pv,Iwasaki:1976cn,Chao:1980dv,Ader:1981db}.
Subsequently, a broad range of approaches have been proposed to explore and understand the nature of these tetraquarks. These include constituent quark models~\cite{Badalian:1985es,Berezhnoy:2011xn,Wu:2016vtq,Barnea:2006sd,Liu:2019zuc,Wang:2019rdo,Bedolla:2019zwg,Deng:2020iqw,Jin:2020jfc,Lu:2020cns,Zhang:2022qtp}, QCD sum rules~\cite{Chen:2016jxd,Wang:2017jtz,Wang:2018poa,Zhang:2020xtb,Wang:2020dlo,Wang:2020ols,Albuquerque:2020hio,Wan:2020fsk,Yang:2020wkh,Wang:2021mma,Wu:2022qwd,Agaev:2023wua,Agaev:2023gaq}, diffusion Monte Carlo simulations~\cite{Bai:2016int,Gordillo:2020sgc}, lattice QCD calculations~\cite{Hughes:2017xie,Meng:2024czd,Li:2025vbd,Li:2025ftn}, and studies based on partial wave analyses~\cite{Liang:2021fzr}, the color evaporation model~\cite{Carvalho:2015nqf,Abreu:2023wwg}, and the Bethe–Salpeter framework~\cite{Li:2021ygk,Ke:2021iyh,Heupel:2012ua}. Related extensions to other fully heavy tetraquark systems can be found in Refs.~\cite{Esposito:2018cwh,Agaev:2025wyf,Wang:2025apq,Xia:2025mgk}.
On the phenomenological side, the production of fully heavy tetraquarks in various collision environments—such as $pp$, $pA$, $AA$, $ep$, $e^+ e^-$, and $\gamma \gamma$—has been considered in Refs.~\cite{Karliner:2016zzc,Berezhnoy:2011xy,Carvalho:2015nqf,Esposito:2018cwh,Wang:2020gmd,Maciula:2020wri,Feng:2020riv,Zhang:2020hoh,Zhu:2020xni,Feng:2020qee,Goncalves:2021ytq,Feng:2023agq,Abreu:2023wwg,Feng:2023ghc,Celiberto:2024mab,Bai:2024ezn,Belov:2024qyi,Bai:2024flh,Liang:2025wbt,Ma:2025ryo,Celiberto:2025dfe,Celiberto:2025ziy,Wang:2025hex,Celiberto:2025vra}. Studies on the electromagnetic and hadronic decays of fully tetraquarks can be found in Refs.~\cite{Biloshytskyi:2022dmo,Sang:2023ncm,Chen:2024orv,Becchi:2020uvq,Zhang:2023ffe,Wang:2023kir,Chen:2022sbf,Lu:2025lyu,Liu:2025mxv}.

In this work, we investigate the exclusive decays of fully tetraquarks into light mesons, specifically $T_{4c,b} \to \pi^+\pi^-$ and $T_{4c,b} \to K^+K^-$. These exclusive decay processes bear notable similarities to the hadronic exclusive decays of charmonia. The first calculation of such exclusive decays in charmonia, exemplified by the decay $J/\psi \to p\bar{p}$, was performed in Ref.~\cite{Brodsky:1981kj}. Subsequently, various decays of $S$- and $P$-wave charmonia have been extensively studied in numerous publications (see, for example, Refs.~\cite{Chernyak:1983ej,Brambilla:2010cs} and references therein).
The factorization approach used in hard exclusive processes for charmonia decays into light hadrons can also be extended to the processes $T_{4c,b} \to \pi^+\pi^-$ and $T_{4c,b} \to K^+K^-$, which we aim to explore in this work.

The paper is organized as follows. In Section~\ref{section-Theoretical-Framework}, we present the theoretical framework, including the factorization formula for the exclusive decay of the $T_{4c}$ into light hadrons $\pi^+\pi^-$ within the context of NRQCD and collinear QCD factorization for hard exclusive processes. We also outline the matching procedure for calculating the hard kernels. 
Additionally, we describe the computational techniques and present the analytical expressions for the hard kernels, along with the master formula for the decay amplitudes, which can be directly utilized for making predictions. 
Section~\ref{section-Phenomenological-Exploration} is dedicated to phenomenological predictions and discussions, while Section~\ref{section-Summary-Outlook} provides a summary and outlook. 
In the Appendix, we discuss the underlying ideas of numerical integration using sector decomposition, with the assistance of the Cheng-Wu theorem and the {\tt QHull} program.

\section{Theoretical Framework}
\label{section-Theoretical-Framework}

\subsection{Factorization Formula}
Similar to the process of charmonium decay into baryon-antibaryon pairs~\cite{Brodsky:1981kj}, we consider the exclusive decays of fully heavy tetraquarks into light mesons. In this section, we take the decay $T_{4c} \to \pi^+\pi^-$ as a concrete example; a similar analysis can be applied to the decays $T_{4c} \to K^+K^-$ and $T_{4b} \to \pi^+\pi^-(K^+K^-)$. 
Within the NRQCD factorization framework~\cite{Bodwin:1994jh,Feng:2020riv,Zhang:2020hoh} for $T_{4c}$ and the collinear factorization approach for hard exclusive reactions~\cite{Lepage:1979zb,Lepage:1979za,Lepage:1980fj,Efremov:1978rn,Efremov:1979qk,Duncan:1979ny}, the amplitude for the process $T^{(J)}_{4c}(P) \to \pi^+(Q) + \pi^-(Q')$ can be factorized as:
\begin{equation}
{\cal A}(T^{(J)}_{4c}(P)\to \pi^+(Q) + \pi^-(Q')) = \sqrt{2m_H} \int_0^1 dx \int_0^1 dy \; \Phi_\pi(x) \left[ \frac{\langle {\cal O}^{(J)}_{\bar3\otimes3}\rangle}{4} T^{(J)}_{\bar3\otimes3}(x,y) + \frac{\langle {\cal O}^{(J)}_{6\otimes\bar6}\rangle}{4} T^{(J)}_{6\otimes\bar6}(x,y)  \right] \Phi_\pi^*(y) ,
\label{factorization_formula}
\end{equation}
where $P$, $Q$ and $Q'$ are the momenta of $T_{4c}$, $\pi^+$ and $\pi^-$ respectively, the factor $\sqrt{2m_H}$ is explicitly factored out to indicate the non-relativistic normalization for the hadron $T_{4c}$, with $m_H(=4m_c)$ representing the mass of $T_{4c}$ and $m_c$ denoting the charm quark mass.

The Long Distance Matrix Elements~(LDME) $\langle{\cal O}^{(J)}_{\bar3\otimes3}\rangle$ and $\langle{\cal O}^{(J)}_{6\otimes\bar6}\rangle$ in \eqref{factorization_formula} are defined as:
\begin{equation}
\langle{\cal O}^{(J)}_{\bar3\otimes3}\rangle = \langle 0\vert {\cal O}^{(J)}_{\bar3\times3} \vert T_{4c}\rangle,
\quad {\rm and} \quad  
\langle{\cal O}^{(J)}_{6\otimes\bar6}\rangle = \langle 0\vert {\cal O}^{(J)}_{6\times\bar6} \vert T_{4c}\rangle, 
\end{equation}
where the operators ${\cal O}^{(J)}_{\bar3\times3}$ and ${\cal O}^{(J)}_{6\otimes\bar6}$ are defined as~\cite{Feng:2020riv,Feng:2023agq}:
\begin{eqnarray}
\mathcal{O}_{\bar3\otimes3}^{(0)} =  -\frac{1}{\sqrt{3}}\left[\psi_a^T\left(i \sigma^2\right) \sigma^i \psi_b\right]\left[\chi_c^{\dagger} \sigma^i\left(i \sigma^2\right) \chi_d^*\right] \mathcal{C}_{\bar3\otimes3}^{a b ; c d} ,
&\quad&
\mathcal{O}_{\bar3\otimes3}^{(1)} =  -\frac{i}{\sqrt{2}} \epsilon_H^{i*} \left[\psi_a^T\left(i \sigma^2\right) \sigma^j \psi_b\right]\left[\chi_c^{\dagger} \sigma^k\left(i \sigma^2\right) \chi_d^*\right] \epsilon^{i j k} \mathcal{C}_{\bar3\otimes3}^{a b ; c d} ,
\nonumber\\
\mathcal{O}_{\bar3\otimes3}^{(2)} =  \frac{1}{2} \epsilon_H^{ij*} \left[\psi_a^T\left(i \sigma^2\right) \sigma^k \psi_b\right]\left[\chi_c^{\dagger} \sigma^l \left(i \sigma^2\right) \chi_d^*\right] \Gamma^{i j ; k l} \mathcal{C}_{\bar3\otimes3}^{a b ; c d} ,
&\quad&
\mathcal{O}_{6\otimes\bar6}^{(0)} =  \left[\psi_a^T\left(i \sigma^2\right) \psi_b\right]\left[\chi_c^{\dagger}\left(i \sigma^2\right) \chi_d^*\right] \mathcal{C}_{6\otimes\bar6}^{a b ; c d} ,
\end{eqnarray}
where $\psi$ and $\chi^\dag$ are Pauli spinor fields that annihilate a heavy quark and anti-quark, respectively. The symbol $\sigma^i$ denotes the 2-dimensional Pauli matrix, $\epsilon^i_H$ and $\epsilon^{ij}_H$ refer to the polarization vector and tensor of $T_{4c}$ with quantum numbers ${\rm J^{PC}=1^{+-}}$ and ${\rm J^{PC}=2^{++}}$, respectively. The color projectors ${\cal C}$ and symmetric traceless tensor $\Gamma$ are defined as:
\begin{eqnarray}
\mathcal{C}_{\bar3\otimes3}^{a b ; c d} \equiv \frac{1}{2 \sqrt{3}}\left(\delta^{a c} \delta^{b d}-\delta^{a d} \delta^{b c}\right), 
\quad
\mathcal{C}_{6\otimes\bar6}^{a b ; c d} \equiv \frac{1}{2 \sqrt{6}}\left(\delta^{a c} \delta^{b d}+\delta^{a d} \delta^{b c}\right),
\quad
\Gamma^{ij;kl} = \delta^{ik}\delta^{jl} + \delta^{il}\delta^{jk} - \frac{2}{3}\delta^{ij}\delta^{kl}
\end{eqnarray}
with $i,j,k$ denoting the Cartesian indices and $a,b,c,d$ representing the color indices. 

The function $\Phi_\pi(x)$ in Eq.~\eqref{factorization_formula} represents the non-perturbative yet universal leading-twist pion Light-Cone Distribution Amplitude (LCDA). This amplitude indicates the probability of finding the valence $u$ and $\bar{d}$ quarks inside the $\pi^+$ meson, carrying fractional momenta $x$ and $\bar{x} \equiv 1 - x$, respectively. The leading-twist pion LCDA is defined as~\cite{Chernyak:1983ej,Braun:2000kw}:
\begin{equation}
\Phi_\pi \left(x, \mu_F\right)=\int \frac{d z^{-}}{2 \pi i} e^{i z^{-} x P^{+}}\langle 0\vert \bar{d}(0) \gamma^{+} \gamma_5 \mathcal{W}(0, z^{-}) u(z^{-})\vert\pi^{+}(P)\rangle,
 \label{pion_LCDA_definition}
\end{equation}
where $P$ is the momentum of the pion and ${\cal W}(0,z^-)$ denotes the light-like gauge link that ensures gauge invariance. By renormalizing the nonlocal operator in \eqref{pion_LCDA_definition}, one recovers the well-known Efremov-Radyushkin-Brodsky-Lepage (ERBL) evolution equation~\cite{Lepage:1980fj,Efremov:1979qk}:
\begin{equation}
\frac{d \Phi_\pi\left(x, \mu_F\right)}{d \ln \mu_F^2} = \int_0^1 dy \; V(x, y) \; \Phi_\pi\left(y, \mu_F\right).
\end{equation}

Finally, the hard kernels $T^{(J)}_{\bar3\otimes3}(x,y)$ and $T^{(J)}_{6\otimes\bar6}(x,y)$ are insensitive to the internal structure of the pion and can be calculated perturbatively. The computation of these hard kernels is one of the primary aims of this work.

\subsection{Perturbative Matching}

\begin{figure}[th]
\includegraphics[width=0.45\textwidth]{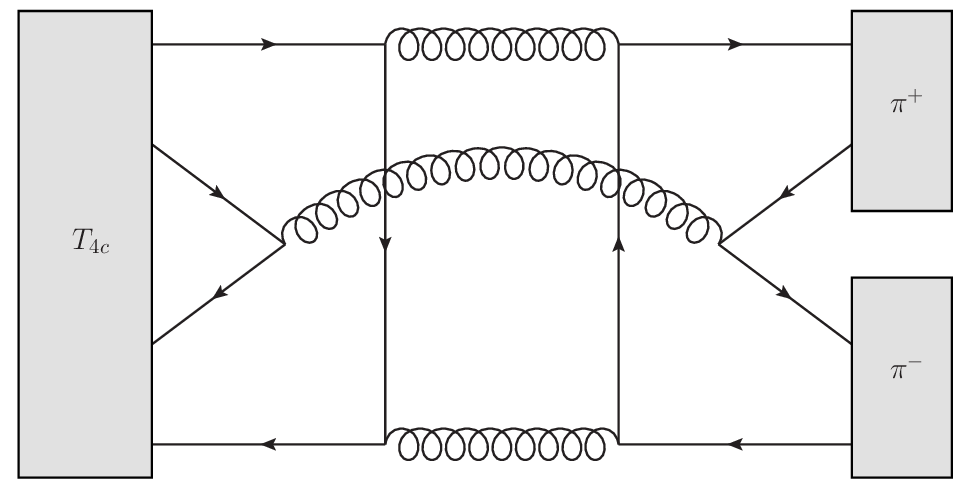}
\caption{A typical Feynman diagram for $T_{4c} \to {\pi^+\pi^-}$.\label{feynman-diagram}}
\end{figure}

By virtue of the factorization theorem, the hard kernels $T_{\bar3\otimes3}^{(J)}(x,y)$ and $T_{6\otimes\bar6}^{(J)}(x,y)$ are insensitive to the internal structure of the hadrons $T_{4c}$ and $\pi^+\pi^-$. Therefore, these hard kernels can be deduced by replacing the hadrons with free partonic states that have the same quantum numbers as the incoming or outgoing hadrons. A typical leading-order Feynman diagram for the process $T_{4c} \to \pi^+\pi^-$ is depicted in FIG.~\ref{feynman-diagram}.

To be more specific, for the incoming fully heavy tetraquark $T_{4c}$, we replace the hadronic state $\vert T_{4c}\rangle$ with the free partonic state $\vert T_{4c} \rangle^{\tt free}$, which consists of free heavy quarks and anti-quarks, represented as $cc\bar c\bar c$. In this configuration, all quarks and anti-quarks share the same momentum $p=\frac{1}{4}P$, where $P$ is the momentum of $T_{4c}$ and satisfies $p^2=m_c^2$. 
For a specific Feynman diagram, we obtain four Dirac spinors associated with the four incoming free quarks or anti-quarks, represented as ${u}_i^a u_j^b \bar{v}_k^c \bar{v}_l^d$. To project the same quantum numbers as the incoming state $\vert T_{4c} \rangle$, we can perform the following replacement for the $\bar{3}\otimes3$ case (it is similar for the $6\otimes\bar{6}$ case):
\begin{eqnarray}
{\color{blue}u}_i^a u_j^b \bar{\color{blue}v}_k^c \bar{v}_l^d \to ({\color{blue}u} u^T)_{ij} (\bar{\color{blue}v}^T\bar{v})_{lk} {\cal {\cal C}}^{ab ;c d}_{\bar3\otimes3}
= ({\color{blue}u} \bar{v} {\rm C})_{ij} ({\rm C}{\color{blue}u}\bar{v})_{lk} {\cal {\cal C}}^{ab ;c d}_{\bar3\otimes3} = (\Pi_s {\rm C})_{ij} ({\rm C} \Pi_{s'})_{lk} {\cal C}^{ab ;c d}_{\bar3\otimes3}
\label{T4Q-replacement}
\end{eqnarray}
where ${\rm C}$ is the charge conjugation matrix, and the diquark $cc$ and di-antiquark $\bar{c}\bar{c}$ have spin $s$ and $s'$, respectively. The spin $s$ and $s'$ can be $0$ for the spin singlet or $1$ for the spin triplet. The corresponding spin projectors are given by~\cite{Petrelli:1997ge}:
\begin{eqnarray}
\Pi_0 = \frac{\gamma^5(\not\!p-m_c)}{2\sqrt{2}m_c}, \quad\quad \Pi_1 = \frac{\gamma^\mu(\not\!p-m_c)}{2\sqrt{2}m_c},
\end{eqnarray}
Note that we adopt non-relativistic normalization for the incoming free quarks or anti-quarks.

Similarly, by replacing the hadronic state $\vert T_{4c}\rangle$ with the partonic state $\vert T_{4c}\rangle^{\text{free}}$, we can explicitly compute the corresponding LDMEs. We express them as follows:
\begin{equation}
\sqrt{2m_H}\langle 0 \vert {\cal O}^{(J)}_{\bar3\otimes3} \vert T_{4c}\rangle \to \langle 0 \vert {\cal O}^{(J)}_{\bar3\otimes3} \vert T_{4c}\rangle^{\tt free} = 4 \epsilon^{(J)}_H, 
\quad 
\sqrt{2m_H} \langle 0 \vert {\cal O}^{(0)}_{6\otimes\bar6} \vert T_{4c}\rangle \to \langle 0 \vert {\cal O}^{(0)}_{6\otimes\bar6} \vert T_{4c}\rangle^{\tt free} = 4.
\end{equation}
where $\epsilon^{(0)}_H=1$, $\epsilon^{(1)}_H$ and $\epsilon^{(2)}_H$ represent the polarization vector of the tetraquark $T^{(1)}_{4c}$ and polarization tensor of the tetraquark $T^{(2)}_{4c}$, respectively.

Furthermore, for the spin $s=s'=1$, one needs to combine both spin angular momenta to obtain the total angular momentum $j=0,1,2$, corresponding to the tetraquark $T_{4c}$ with quantum numbers ${\rm J^{PC}=0^{++},1^{+-}, 2^{++}}$, respectively. Utilizing Clebsch–Gordan coefficients, we can express the contraction $J_j^{\mu\nu} \Pi_1^\mu \Pi_1^\nu$, which projects the final angular momentum $j$~\cite{Braaten:2002fi}:
\begin{eqnarray}
J_0^{\mu \nu}= \frac{1}{\sqrt{3}} \eta^{\mu \nu}, 
\,
J_1^{\mu \nu} = -\frac{i}{\sqrt{2 p^2}} \varepsilon^{\mu \nu \rho \sigma} \epsilon_\rho p_\sigma, 
\,
J_2^{\mu \nu} = \epsilon_{\rho \sigma}\left\{\frac{1}{2}\left[\eta^{\mu \rho} \eta^{\nu \sigma}+\eta^{\mu \sigma} \eta^{\nu \rho}\right] 
 -\frac{1}{3} \eta^{\mu \nu} \eta^{\rho \sigma}\right\},
\end{eqnarray}
where $\eta^{\mu \nu}=-g^{\mu \nu}+p^\mu p^\nu/p^2$, $\epsilon_\rho$ is the polarization vector of $T^{(1)}_{4c}$, and $\epsilon_{\rho\sigma}$ is the polarization tensor of $T^{(2)}_{4c}$.

Next, let’s turn our attention to the light hadron pion. It is straightforward to project the correct quantum numbers for the pion using the following replacement:
\begin{equation}
v^a_i(\bar xQ) \bar{u}^b_j(xQ) \to \frac{1}{2} (\gamma_5 Q\!\!\!\!/)_{ij} \delta^{ab}, \mbox{ which is compatible with } \Phi_\pi(x,\mu_F) \xrightarrow{\mu_F \to \infty} \frac{f_\pi}{2\sqrt{2N_c}} 6x\bar{x}, 
\label{pion-replacement}
\end{equation}
where $Q$ is the momentun of $\pi$ and $f_\pi$ is the decay constant for the pion.

We utilize the program {\tt HepLib}~\cite{Feng:2021kha,Feng:2023hxy} to perform the operations described above. {\tt HepLib} is a {\tt C++} library designed for high-energy physics calculations, which interfaces with {\tt QGRAF}~\cite{Nogueira:1991ex} to generate Feynman diagrams and amplitudes. In this context, we generate the partonic process: $c(p)c(p)\bar{c}(p)\bar{c}(p) \to u(xQ)\bar{d}(\bar xQ) + \bar u(yQ') d(\bar yQ')$, resulting in the generation of 150 diagrams at leading order. This number is much more than that found for the process $J/\psi \to p\bar{p}$, which involves only a few Feynman diagrams.

Each diagram consists of four fermion lines and at least three gluon lines. The general structure for each diagram can be expressed as:
\begin{equation}
	{\cal A} = {u}_{i'}^{a'}(p) u_{j'}^{b'}(p) \bar{v}_{k'}^{c'}(p) \bar{v}_{l'}^{d'}(p) \; {\cal H}^{i'j'k'l'ijkl}_{a'b'c'd'abcd} \; 
    v_i^a(\bar xQ) \bar{u}_j^b(xQ) \, u_k^c(yQ') \bar{v}_l^d(\bar yQ') ,
\end{equation}
using the method outlined in this section, we replace ${u}_{i'}^{a'} u_{j'}^{b'} \bar{v}_{k'}^{c'} \bar{v}_{l'}^{d'}$ with the expression from  \eqref{T4Q-replacement}, and $v_i^a \bar{u}_j^b$ and $u_k^c \bar{v}_l^{d}$ with \eqref{pion-replacement} and its charge-conjugated version, respectively. 
After contracting all Dirac-$\gamma$ matrix indices, the two heavy quark lines and two light quark lines yield Dirac traces. Following the computation of these traces and contractions (with {\tt HepLib} internally invoking the {\tt FORM}~\cite{Vermaseren:2000nd} program to perform operations automatically), we obtain the final expressions for the hard kernels $T_{\bar{3}\otimes3}^{(J)}$ and $T_{6\otimes\bar{6}}^{(J)}$ as follows:
\begin{eqnarray}
T^{(0)}_{\bar3\otimes3} &=&
(\alpha_s \pi)^3 (
2048 x^5 y^3-3072 x^5 y^2+1408 x^5 y-192 x^5-512 x^4 y^4-4096 x^4 y^3 +5888 x^4 y^2
\label{T1} \\ &&
-2240 x^4 y+256 x^4+2048 x^3 y^5-4096 x^3 y^4+8320 x^3 y^3 -6848 x^3 y^2+1684 x^3 y
\nonumber \\ &&
-106 x^3-3072 x^2 y^5+5888 x^2 y^4-6848 x^2 y^3 +4072 x^2 y^2-742 x^2 y+17 x^2+1408 x y^5
\nonumber \\ &&
-2240 x y^4+1684 x y^3-742 x y^2
+114 x y-2 x-192 y^5+256 y^4-106 y^3+17 y^2-2 y
)
\nonumber \\ &&
\div [9 m_c^6 x (2 x-1)^2 \bar{x} y (2 y-1)^2 \bar{y} (4 x y-x-y) (4 \bar{x}\bar{y}-\bar{x}-\bar{y})]
, \nonumber \\
T^{(0)}_{6\otimes\bar 6} &=&
(\alpha_s \pi)^3 (
2048 x^5 y^3-3072 x^5 y^2+1408 x^5 y-192 x^5-512 x^4 y^4-4096 x^4 y^3+5888 x^4 y^2
\label{T2} \\ &&
-2240 x^4 y+256 x^4+2048 x^3 y^5-4096 x^3 y^4+8320 x^3 y^3-6848 x^3 y^2+1620 x^3 y
\nonumber \\ &&
-74 x^3-3072 x^2 y^5+5888 x^2 y^4-6848 x^2 y^3+4200 x^2 y^2-774 x^2 y+x^2+1408 x y^5
\nonumber \\ &&
-2240 x y^4+1620 x y^3-774 x y^2+146 x y-2 x-192 y^5+256 y^4-74 y^3+y^2-2 y
)
\nonumber \\ &&
\div [3\sqrt{6} m_c^6 x (2 x-1)^2 \bar{x} y (2 y-1)^2 \bar{y} (4 x y-x-y) (4 \bar{x}\bar{y}-\bar{x}-\bar{y})]
, \nonumber \\
T^{(2)}_{\bar 3\otimes 3} &=&
-(\alpha_s \pi)^3 (4 m_c^2 g^{\mu\nu} + 3 K^\mu K^\nu) \epsilon_{\mu\nu} (
5120 x^5 y^3-7680 x^5 y^2+3520 x^5 y-480 x^5-512 x^4 y^4-11776 x^4 y^3
\label{T3} \\ &&
+18944 x^4 y^2-9056 x^4 y+1264 x^4+5120 x^3 y^5-11776 x^3 y^4+20800 x^3 y^3-20960 x^3 y^2+8956 x^3 y
\nonumber \\ &&
-1198 x^3-7680 x^2 y^5+18944 x^2 y^4-20960 x^2 y^3+13168 x^2 y^2-4282 x^2 y+485 x^2+3520 x y^5
\nonumber \\ &&
-9056 x y^4+8956 x y^3-4282 x y^2+972 x y-71 x-480 y^5+1264 y^4-1198 y^3+485 y^2-71 y
)
\nonumber \\ &&
\div [36\sqrt{3} m_c^8 x (2 x-1)^2 \bar{x} y (2 y-1)^2 \bar{y} (4 x y-x-y) (4 \bar{x}\bar{y}-\bar{x}-\bar{y})]
, \nonumber
\end{eqnarray}

the other contributions are zero, {\it i.e.},
\begin{eqnarray}
T^{(1)}_{\bar 3\otimes 3} = T^{(1)}_{6\otimes\bar 6} = T^{(2)}_{6\otimes \bar 6} = 0 .
\end{eqnarray}

Note that the $i\varepsilon$-prescription is necessary for the denominators to ensure the correct identification of the branching cuts. The explicit forms of the $i\varepsilon$ terms are as follows:
\begin{eqnarray}
(2x-1)^2 \to (2x-1+i\varepsilon)(2x-1-i\varepsilon) = -(2x-1+i\varepsilon)(2\bar x-1+i\varepsilon) \,,
\\
4xy-x-y \to 4xy-x-y + i\varepsilon, \quad  4\bar x\bar y-\bar x-\bar y \to 4\bar x\bar y-\bar x-\bar y + i\varepsilon \,.
\end{eqnarray}
It is easy to verify that the hard kernels possess the following symmetry:
\begin{equation}
T_{\bar3\otimes3}^{(J)}(x,y)=T_{\bar3\otimes3}^{(J)}(1-x,1-y) , \quad
T_{6\otimes\bar6}^{(J)}(x,y)=T_{6\otimes\bar6}^{(J)}(1-x,1-y) .
\end{equation}
Note that this symmetry also holds in the case with the $i\varepsilon$-prescription.

\subsection{Master Formula for Amplitude}

The leading-twist pion LCDA can be conveniently expanded in the Gegenbauer polynomial basis:
\begin{equation}
\Phi_\pi(x,\mu_F) = \frac{f_\pi}{2\sqrt{2N_c}} {\sum_{n = 0}}' a_n(\mu_F) \psi_n(x), \mbox{ with } \psi_n(x) \equiv 6x\bar{x} C_n^{3/2}(2x-1)
\label{phi-expansion}
\end{equation}
where the pion decay constant $f_\pi = 0.131$ GeV, and $\sum'$ signifies the sum over even integers.
All the nonperturbative dynamics are encoded in the Gegenbauer moments $a_n(\mu_F)$.
Those non-perturbative parameters $a_n(\mu_F)$ at the factorization scale $\mu_F$ can be evolved from an initial scale $\mu_0$ using the one-loop evolution equation:
\begin{equation}
a_n^{1-{\rm loop}}(\mu_F)=a_n(\mu_0)\left[\frac{\alpha_s(\mu_F^2)}{\alpha_s(\mu_0^2)}\right]^{\gamma(n)} \mbox{ with } \gamma(n) \equiv \frac{\gamma_0(n)}{2 b_0} ,
\end{equation}
where the standard $\beta$-function coefficient $b_0$ is given by $b_0 = (11N_c-2n_f)/3$, and the one-loop anomalous dimensions $\gamma_0(n)$ are expressed as:
\begin{equation}
\gamma_0(n)=2 C_F\left[4 \sum_{i=1}^{n+1} \frac{1}{i}-3-\frac{2}{(n+1)(n+2)}\right].
\end{equation}

For the kaon, one also needs to sum over the odd integers in addition to the even integers, {\it i.e.},
\begin{equation}
\Phi_K(x,\mu_F) = \frac{f_K}{2\sqrt{2N_c}} {\sum_{n = 0}} a_n(\mu_F) \psi_n(x),
\end{equation}
where the kaon decay constant is given by $f_K = 0.160{\rm GeV}$.

By substituting the expansion in $\eqref{phi-expansion}$ into the factorization fomula in $\eqref{factorization_formula}$, we obtain the following integrated master formula for the amplitude of the process $T^{(0)}_{4c}\to\pi^+\pi^-$:
\begin{equation}
{\cal A}(T_{4c}^{(0)} \to \pi^+\pi^-) = {\sqrt{2m_H} \frac{(\alpha_s \pi)^3}{m_c^6} f^2_\pi } \left[ 
\frac{\langle {\cal O}^{(0)}_{\bar3\otimes3}\rangle}{4} \sum_{m,n} {\cal D}^{(0)}_{mn}\, a_m(\mu_F) a_n(\mu_F)
+ \frac{\langle {\cal O}^{(0)}_{6\otimes\bar6}\rangle}{4} \sum_{m,n} {\cal F}^{(0)}_{mn}\, a_m(\mu_F) a_n(\mu_F) \right]
\end{equation}
where the integrated coefficients ${\cal D}^{(0)}_{mn}$ and ${\cal F}^{(0)}_{mn}$ are introducted as:
\begin{equation}
\frac{(\alpha_s \pi)^3}{m_c^6}{\cal D}^{(0)}_{mn} = \int_0^1 dx \int_0^1 dy \, \frac{\psi_m(x)}{2\sqrt{2N_c}} \, T_{\bar3\otimes3}^{(J)}(x,y) \, \frac{\psi_n(y)}{2\sqrt{2N_c}} ,\;
\frac{(\alpha_s \pi)^3}{m_c^6}{\cal F}^{(0)}_{mn} = \int_0^1 dx \int_0^1 dy \, \frac{\psi_m(x)}{2\sqrt{2N_c}} \, T_{6\otimes\bar6}^{(J)}(x,y) \, \frac{\psi_n(y)}{2\sqrt{2N_c}}
\label{conv-1-def}
\end{equation}

For the process $T_{4c}^{(2)} \to \pi^+\pi^-$, we have:
\begin{equation}
{\cal A}(T_{4c}^{(2)} \to \pi^+(P)+\pi^-(K)) = {\sqrt{2m_H} \frac{(\alpha_s \pi)^3}{m_c^8} f^2_\pi } (4 m_c^2 g^{\mu\nu} + 3 P^\mu P^\nu) \epsilon_{\mu\nu}
\left[ \frac{\langle {\cal O}^{(2)}_{\bar3\otimes3}\rangle}{4} \sum_{m,n} {\cal D}^{(2)}_{mn}\, a_m(\mu_F) a_n(\mu_F) \right],
\end{equation}
where ${\cal D}^{(2)}_{mn}$ is defined as:
\begin{equation}
\frac{(\alpha_s \pi)^3}{m_c^8} (4 m_c^2 g^{\mu\nu} + 3 P^\mu P^\nu) \epsilon_{\mu\nu} {\cal D}^{(2)}_{mn} = \int_0^1 dx \int_0^1 dy \, \frac{\psi_m(x)}{2\sqrt{2N_c}} \, T_{\bar3\otimes3}^{(2)}(x,y) \, \frac{\psi_n(y)}{2\sqrt{2N_c}} ,
\label{conv-2-def}
\end{equation}
and the polarization tensor $\epsilon_{\mu\nu}$ satisfies the following summation:
\begin{equation}
\sum_{\rm pol.} \vert (4 m_c^2 g^{\mu\nu} + 3 P^\mu P^\nu) \epsilon_{\mu\nu} \vert^2 = 96 m_c^4 .
\end{equation}

Note that the convolutions in \eqref{conv-1-def} and \eqref{conv-2-def} cannot be integrated straightforwardly using {\tt PolyLogTools}~\cite{Duhr:2019tlz} package due to the presence of $i\varepsilon$ in the denominators. 
While one might attempt to transorm the $i\varepsilon$ into $x$ as $x+i\varepsilon$ and $\bar{x}$ as $\bar{x}+i\varepsilon$, this leads to contradictions with the constraint $x+\bar{x}=1$. 
Therefore, we employ the sector docomposition method~\cite{Binoth:2003ak,Heinrich:2008si}, with the assistance of the Cheng-Wu theorem~\cite{Cheng:1987ga,Smirnov:2006ry}, to numerically integrate these convolutions. The details can be found in Appendix~\ref{sec:appendix}. This method can be easily extended to other process, such as $T_{4c,b} \to p\bar{p}$, and to various types of integration of phase-space in high-energy physics.

The convolution results for a few leading values of $m$ and $n$ are tabulated in the TABLE~\ref{conv-table}. As observed, when $m+n$ is odd, the convolutions yield zero, it can be easily understood since $T_{\bar3\otimes3}^{(0)}(x,y)=T_{\bar3\otimes3}^{(0)}(1-x,1-y)$ and $\psi_m(x)=(-1)^m \psi_m(1-x)$. A similar reasoning applies to the cases of $T^{(0)}_{6\otimes\bar6}(x,y)$ and $T^{(2)}_{\bar3\otimes3}(x,y)$.

\begin{table}[!ht]
\centering
\setlength{\tabcolsep}{2pt}
\begin{tabular}{c|c|c|c||c|c|c|c}
\hline\hline
$(m,n)$ & ${\cal D}^{(0)}_{mn}$  & ${\cal F}^{(0)}_{mn}$ & ${\cal D}^{(2)}_{mn}$ & 
$(m,n)$ & ${\cal D}^{(0)}_{mn}$  & ${\cal F}^{(0)}_{mn}$ & ${\cal D}^{(2)}_{mn}$ \\
\hline\hline
(0, 0) & $0.9718 - 0.01863 i$ & $1.705 - 0.6430 i$ & $0.2361 - 0.3257 i$ &
 (3, 3) & $16.10 - 56.55 i$ & $17.42 - 74.67 i$ & $2.094 - 9.631 i$\\
(0, 2) & $-5.257 + 3.416 i$ & $-7.071 + 7.514 i$ & $-0.9829 + 1.351 i$ &
 (3, 5) & $-28.31 + 120.2 i$ & $-32.41 + 156.7 i$ & $-3.922 + 19.56 i$\\
(0, 4) & $14.19 - 17.05 i$ & $17.55 - 26.88 i$ & $2.200 - 4.052 i$ &
 (3, 7) & $39.95 - 215.7 i$ & $45.85 - 275.9 i$ & $5.570 - 34.00 i$\\
(0, 6) & $-29.96 + 49.75 i$ & $-36.66 + 71.74 i$ & $-4.470 + 9.750 i$ &
 (4, 4) & $49.99 - 105.2 i$ & $63.93 - 144.9 i$ & $8.035 - 19.10 i$\\
(0, 8) & $52.72 - 112.1 i$ & $64.26 - 152.5 i$ & $7.651 - 19.89 i$ &
 (4, 6) & $-83.63 + 213.0 i$ & $-106.9 + 283.0 i$ & $-13.12 + 36.14 i$\\
(1, 1) & $2.518 - 5.502 i$ & $2.405 - 9.078 i$ & $0.3080 - 1.184 i$ &
 (4, 8) & $133.1 - 387.6 i$ & $166.8 - 504.4 i$ & $20.19 - 63.10 i$\\
(1, 3) & $-5.745 + 18.53 i$ & $-6.701 + 26.33 i$ & $-0.8781 + 3.424 i$ &
 (5, 5) & $53.46 - 242.6 i$ & $64.13 - 312.3 i$ & $7.292 - 38.39 i$\\
(1, 5) & $9.728 - 42.11 i$ & $12.27 - 57.26 i$ & $1.445 - 7.168 i$ &
 (5, 7) & $-79.10 + 419.2 i$ & $-94.08 + 531.8 i$ & $-11.00 + 64.95 i$\\
(1, 7) & $-12.39 + 79.14 i$ & $-16.83 + 102.9 i$ & $-2.088 + 12.74 i$ &
 (6, 6) & $123.7 - 383.6 i$ & $158.1 - 497.3 i$ & $19.21 - 62.40 i$\\
(2, 2) & $13.36 - 14.70 i$ & $19.33 - 24.90 i$ & $2.372 - 3.785 i$ &
 (6, 8) & $-181.7 + 643.0 i$ & $-228.4 + 823.8 i$ & $-27.77 + 101.8 i$\\
(2, 4) & $-28.97 + 45.51 i$ & $-38.05 + 65.99 i$ & $-4.777 + 9.226 i$ &
 (7, 7) & $123.5 - 702.5 i$ & $144.3 - 885.1 i$ & $17.00 - 107.4 i$\\
(2, 6) & $54.70 - 108.9 i$ & $70.24 - 148.6 i$ & $8.450 - 19.63 i$ &
 (8, 8) & $246.5 - 1012. i$ & $310.2 - 1287. i$ & $37.49 - 157.3 i$\\
(2, 8) & $-93.35 + 219.9 i$ & $-115.5 + 290.2 i$ & $-13.94 + 37.18 i$ &
 $*$ & $0$ & $0$ & $0$ \\
\hline\hline
\end{tabular}
\caption{Numerical values for ${\cal D}^{(0)}_{mn}$, ${\cal F}^{(0)}_{mn}$ and ${\cal D}^{(2)}_{mn}$, where the $*$ case refers that $(m+n)$ is odd.\label{conv-table}}
\end{table}

\section{Phenomenological Exploration}  
\label{section-Phenomenological-Exploration}
Now, we are ready to present the decay widths for the process $T^{(J)}_{4c,b} \to \pi^+\pi^- (K^+K^-)$ utilizing the calculated amplitude ${\cal A}(T^{(J)}_{4c,b} \to \pi^+\pi^- (K^+K^-))$, {\it i.e.},
\begin{equation}
\Gamma(T^{(J)}_{4c,b} \to \pi^+\pi^-(K^+K^-)) = \frac{1}{2m_H} \frac{\Phi_2}{2J+1} \vert {\cal A}(T^{(J)}_{4c,b}\to \pi^+\pi^-(K^+K^-))\vert^2
\end{equation}
where $\Phi_2 = 1/(8\pi)$ is the massless two-body phase space.

Before making phenomenological predictions, we need to fix the various input parameters. We take the charm quark mass to be $m_c=1.5~ {\rm GeV}$. The QCD running coupling constant at the renormailizaiton scale $\mu_R$ is evaluated with the aid of the package {\tt RunDec}~\cite{Chetyrkin:2000yt}. Furthermore, we also take $\mu_R = \mu_F = \mu$.

Similar to the previous work~\cite{Sang:2023ncm}, the LDMEs for $T_{4c}$ are taken or extracted from Refs.~\cite{Lu:2020cns} and \cite{liu:2020eha}, which we will refer to as {\tt Model I} and {\tt Model II}, repsectively. These values for the LDMEs of $T_{4c}$ are tabulaed in TABLE~\ref{LDME-table}.
\begin{table}[!ht]
\centering
\setlength{\tabcolsep}{13pt}
\begin{tabular}{c||ccc|c|c}
\hline\hline
LDMEs & $\vert\langle{\cal O}_{\bar3\otimes3}^{(0)}\rangle\vert^2$ & $ \langle{\cal O}_{\bar3\otimes3}^{(0)}\rangle \langle{\cal O}_{6\otimes\bar6}^{(0)}\rangle $ & $\vert\langle{\cal O}_{6 \otimes \bar{6}}^{(0)}\rangle\vert^2$ & $\vert\langle{\cal O}_{\bar{3} \otimes 3}^{(1)}\rangle\vert^2$ & $\vert\langle{\cal O}_{\bar{3} \otimes 3}^{(2)}\rangle\vert^2$ \\
\hline
{\tt Model I}~\cite{Lu:2020cns} & 0.0347 & 0.0211 & 0.0128 & 0.0260 & 0.0144 \\
{\tt Model II}~\cite{liu:2020eha} & 0.0187 & -0.0161 & 0.0139 & 0.0160 & 0.0126 \\
\hline \hline
\end{tabular}
\caption{Numerical values of the LDMEs for $T_{4 c}$ in {\tt Model I}~\cite{Lu:2020cns} and {\tt Model II}~\cite{liu:2020eha}, in unit of $\mathrm{GeV}^9$.\label{LDME-table}}
\end{table}

Similar to Refs. \cite{Chen:2023byr,Chen:2024oem}, the input non-perturbative parameters $a_n(\mu_F)$ for $\pi$ and $K$ are taken from {\tt RQCD}~\cite{RQCD:2019osh} and {\tt LPC}~\cite{LatticeParton:2022zqc} Collaborations, and these values are listed in TABLE~\ref{an-table}.
\begin{table}[!ht]
\centering
\setlength{\tabcolsep}{5pt}
\begin{tabular}{c||cccc}
\hline\hline
$\pi$ & $a_2(2{\rm GeV})$ & $a_4(2{\rm GeV})$  & $a_6(2{\rm GeV})$ & $a_8(2{\rm GeV})$ \\
\hline
 {\tt RQCD}~\cite{RQCD:2019osh} & $0.116^{+0.019}_{-0.020}$ & - & - & - \\
 {\tt LPC}~\cite{LatticeParton:2022zqc} & $ 0.258 \pm 0.087$ & $0.122 \pm 0.056$ & $ 0.068 \pm 0.038$ & - \\
\hline\hline
$K$ & $a_1(2{\rm GeV})$ & $a_2(2{\rm GeV})$  & $a_3(2{\rm GeV})$ & $a_4(2{\rm GeV})$ \\
\hline
 {\tt RQCD}~\cite{RQCD:2019osh} & $-0.0525^{+31}_{-33}$ & $0.106^{+15}_{-16}$ & - & - \\
 {\tt LPC}~\cite{LatticeParton:2022zqc} & $-0.108 \pm 0.014 \pm 0.051$ & $0.170 \pm 0.014 \pm 0.044$ & $-0.043 \pm 0.006 \pm 0.022$ & $0.073 \pm 0.008 \pm 0.021$ \\
\hline\hline
\end{tabular}
\caption{The non-perturbative input parameters $a_n(\mu_F)$ for $\pi$ and $K$ from {\tt RQCD}~\cite{RQCD:2019osh} and {\tt LPC}~\cite{LatticeParton:2022zqc} Collaborations.\label{an-table}}
\end{table}

\begin{figure}[!ht]
\includegraphics[width=0.45\textwidth]{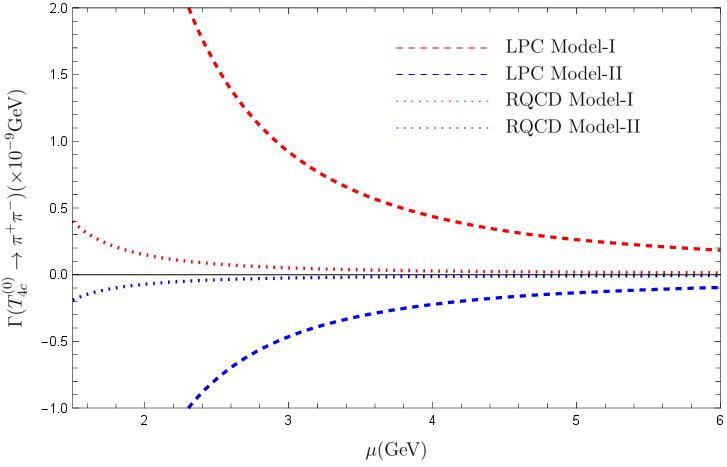}
\includegraphics[width=0.45\textwidth]{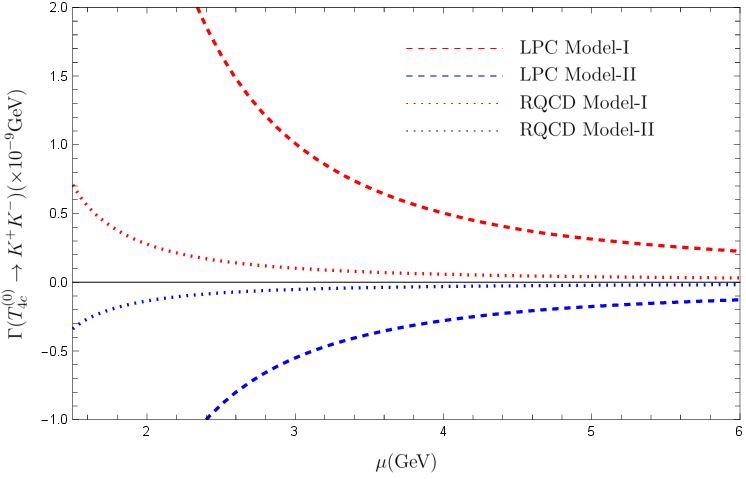}
\includegraphics[width=0.45\textwidth]{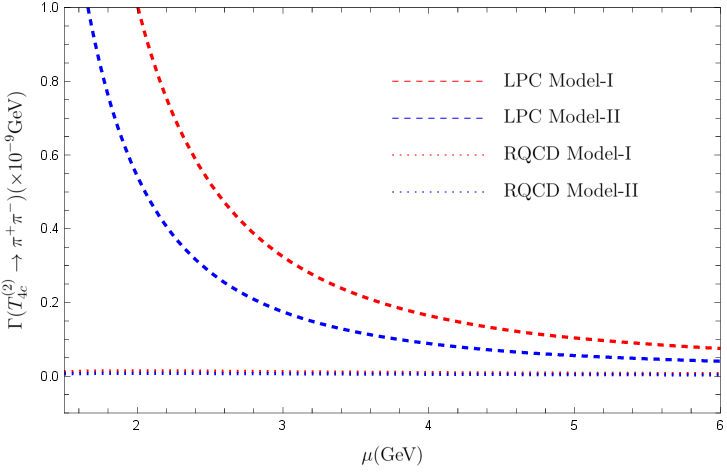}
\includegraphics[width=0.45\textwidth]{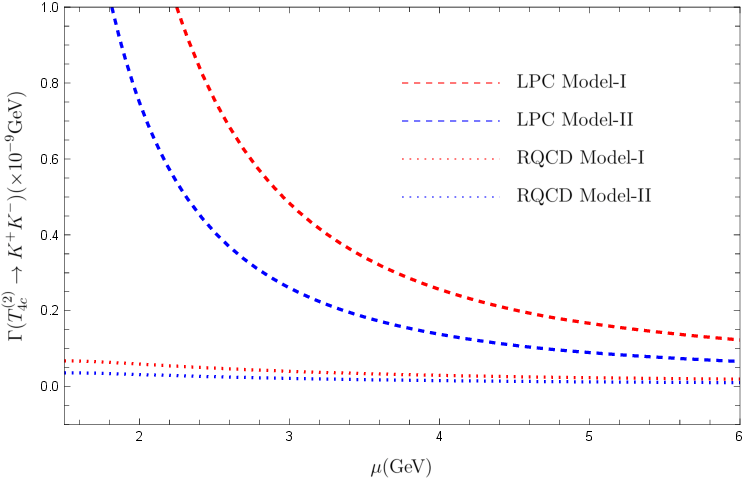}
\caption{Decay widths for $T_{4c} \to \pi^+\pi^-$ and $T_{4c} \to K^+K^-$ as a function of the renormalization/factorization scale $\mu$. The input LDMEs for $T_{4c}$ are taken from {\tt Model I}~\cite{Lu:2020cns} and {\tt Model II}~\cite{liu:2020eha}, while the input non-perturbative parameters $a_n(\mu_F)$ for $\pi$ and $K$ are taken from the {\tt RQCD}~\cite{RQCD:2019osh} and {\tt LPC}~\cite{LatticeParton:2022zqc} Collaborations. \label{T4c-plots}}
\end{figure}

In FIG.~\ref{T4c-plots}, we present the decay widths of $T_{4c} \to \pi^+\pi^-$ and $T_{4c}\to K^+K^-$ as functions of the renormalization/factorization scale $\mu$. As can be seen, the dependence on the factorization/renormalization scale $\mu$ is relatively large at samll values of $\mu$. This dependence is expected to diminish with the inclusion of higher-order corrections. Furthermore, the predicted decay widths using the input parameters from the {\tt LPC} collaboration are significantly larger than those obtained from {\tt RQCD} for both the pion and the kaon. 
It is noteworthy that the predictions for $T^{(0)}_{4c}$ using the LDMEs from {\tt Model II} are negative and unphysical. This is directly attributed to the negative value of the LDME $ \langle{\cal O}_{\bar3\otimes3}^{(0)}\rangle \langle{\cal O}_{6\otimes\bar6}^{(0)}\rangle $ in {\tt Model II} as can be observed in TABLE~\ref{LDME-table}. This finding also indicates that the contribution from the mixing between the $\bar3\otimes3$ and $6\otimes\bar6$ is significant compared to other contributions. 
To illustrate this concretely, we take $T^{(0)}_{4c} \to \pi^+\pi^-$ with the {\tt LPC} input as an example. 
The corresponding decay width at $\mu=2m_c$ is given by:
\begin{equation}
\Gamma(T_{4c}^{(0)} \to \pi^+\pi^-) = \left[\frac{(\pi\alpha_s)^3\, f_\pi^2}{m_c^6}\right]^2 \left\{ 
0.013 \vert\langle{\cal O}_{\bar3\otimes3}^{(0)}\rangle\vert^2 
+0.038 \langle{\cal O}_{\bar3\otimes3}^{(0)}\rangle \langle{\cal O}_{6\otimes\bar 6}^{(0)}\rangle 
+0.028 \vert\langle{\cal O}_{6\otimes\bar 6}^{(0)}\rangle\vert^2 \right\}_{\mu=2m_c}
\end{equation}
where it can be observed that the coefficient of $\langle{\cal O}_{\bar3\otimes3}^{(0)}\rangle \langle{\cal O}_{6\otimes\bar 6}^{(0)}\rangle$ is larger than the coefficients for $\vert\langle{\cal O}_{\bar3\otimes3}^{(0)}\rangle\vert^2$ and $\vert\langle{\cal O}_{6\otimes\bar 6}^{(0)}\rangle\vert^2$.

Furthermore, to predict the braching ratio of the exclusive decays of $T_{4c}$ into pions or kaons, we need to determine the total decay width of $T_{4c}$. Here, we assume that the decay of $T_{4 c}$ is dominated by its decay into double $J/\psi$, allowing us to approximate the total decay width of $T_{4 c}$ as follows:
\begin{equation}
\Gamma_{\rm total}(T_{4 c}) \approx \Gamma(T_{4 c} \to J/\psi J/\psi) \approx 0.12~{\rm GeV},
\end{equation}

To make concrete numerical predictions, we present the detailed decay widths in TABLE~\ref{gamma-br-table}. The decay widths of $T^{(J)}_{4c} \to \pi^+\pi^-$ and $T^{(J)}_{4c} \to K^+K^-$ are around $10^{-9}$ GeV, and the branching ratio is approximately $10^{-8}$, which remains unobservable in current experiments.

\begin{table}[!ht]
\centering
\setlength{\tabcolsep}{12pt}
\begin{tabular}{c||c|c|c|c}
\hline\hline
 & {\tt LPC} + {\tt Model I} & {\tt LPC} + {\tt Model II} & {\tt RQCD} + {\tt Model I} & {\tt RQCD} + {\tt Model II} \\
\hline\hline
$\Gamma(T^{(0)}_{4c} \to \pi^+\pi^-)$ & $9.2\times10^{-10}{\rm GeV}$ & $-4.6\times10^{-10}{\rm GeV}$ & $5.1\times10^{-11}{\rm GeV}$ & $-2.5\times10^{-11}{\rm GeV}$ \\
${\rm Br}(T^{(0)}_{4c} \to \pi^+\pi^-)$ & $7.7\times10^{-9}$ & $-3.9\times10^{-9}$ & $4.2\times10^{-10}$ & $-2.1\times10^{-10}$ \\
\hline
$\Gamma(T^{(0)}_{4c} \to K^+K^-)$ & $1.0\times10^{-9}{\rm GeV}$ & $-5.5\times10^{-10}{\rm GeV}$ & $1.0\times10^{-10}{\rm GeV}$ & $-5.3\times10^{-11}{\rm GeV}$ \\
${\rm Br}(T^{(0)}_{4c} \to K^+K^-)$ & $8.4\times10^{-9}$ & $-4.6\times10^{-9}$ & $8.5\times10^{-10}$ & $-4.4\times10^{-10}$ \\
\hline
$\Gamma(T^{(2)}_{4c} \to \pi^+\pi^-)$ & $3.2\times10^{-9}{\rm GeV}$ & $1.7\times10^{-10}{\rm GeV}$ & $1.2\times10^{-11}{\rm GeV}$ & $6.6\times10^{-12}{\rm GeV}$ \\
${\rm Br}(T^{(2)}_{4c} \to \pi^+\pi^-)$ & $2.7\times10^{-9}$ & $1.5\times10^{-9}$ & $1.0\times10^{-10}$ & $5.5\times10^{-11}$ \\
\hline
$\Gamma(T^{(2)}_{4c} \to K^+K^-)$ & $4.8\times10^{-10}{\rm GeV}$ & $2.6\times10^{-10}{\rm GeV}$ & $4.1\times10^{-11}{\rm GeV}$ & $2.2\times10^{-11}{\rm GeV}$ \\
${\rm Br}(T^{(2)}_{4c} \to K^+K^-)$ & $4.0\times10^{-9}$ & $2.2\times10^{-9}$ & $3.4\times10^{-10}$ & $1.8\times10^{-10}$ \\
\hline\hline
$\Gamma(T^{(0)}_{4b} \to \pi^+\pi^-)$ & $1.7\times10^{-14}{\rm GeV}$ & $-9.2\times10^{-15}{\rm GeV}$ & $1.6\times10^{-15}{\rm GeV}$ & $-9.0\times10^{-16}{\rm GeV}$ \\
$\Gamma(T^{(0)}_{4b} \to K^+K^-)$ & $2.3\times10^{-14}{\rm GeV}$ & $-1.3\times10^{-14}{\rm GeV}$ & $3.8\times10^{-15}{\rm GeV}$ & $-2.2\times10^{-15}{\rm GeV}$ \\
$\Gamma(T^{(2)}_{4b} \to \pi^+\pi^-)$ & $7.7\times10^{-15}{\rm GeV}$ & $4.1\times10^{-15}{\rm GeV}$ & $1.0\times10^{-15}{\rm GeV}$ & $5.4\times10^{-16}{\rm GeV}$ \\
$\Gamma(T^{(2)}_{4b} \to K^+K^-)$ & $1.3\times10^{-14}{\rm GeV}$ & $7.1\times10^{-15}{\rm GeV}$ & $2.8\times10^{-15}{\rm GeV}$ & $1.5\times10^{-15}{\rm GeV}$ \\
\hline\hline
\end{tabular}
\caption{The decay widths and branching ratios for $T_{4c,b} \to \pi^+\pi^- (K^+K^-)$, using input non-perturbative parameters $a_n(\mu)$ from the {\tt RQCD}~\cite{RQCD:2019osh} and {\tt LPC}~\cite{LatticeParton:2022zqc} Collaborations, as well as LDMEs from {\tt Model I}~\cite{Lu:2020cns} and {\tt Model II}~\cite{liu:2020eha}. The renormalization/factorization scale is set to $\mu=2m_c$ for $T_{4c}$ and $\mu=2m_b$ for $T_{4b}$.\label{gamma-br-table}}
\end{table}

Due to the lack of LDMEs $\langle{\cal O}\rangle_{T_{4b}}$ (including $\langle{\cal O}^{(J)}_{\bar3\otimes3}\rangle$ and $\langle{\cal O}^{(J)}_{6\otimes\bar6}\rangle$) of $T_{4b}$, we use a very crude estimate for the LDMEs of $T_{4b}$ as done in Refs. \cite{Feng:2023agq,Sang:2023ncm}, {\it i.e.},

\begin{equation}
\langle {\cal O}\rangle_{T_{4b}} 
\approx \langle {\cal O} \rangle_{T_{4 c}} \times \left(\frac{m_b \alpha_s^b}{m_c \alpha_s^c} \right)^{9/2},
\end{equation}
where $\alpha_s^b$ and $\alpha_s^c$ represent the strong couplings $\alpha_s(m_c v_c) \sim v_c$ and $\alpha_s(m_b v_b) \sim v_b$, respectively, with $v_c$ and $v_b$ denoting the typical velocities of the charm and bottom quarks. 
In this work, we take $m_b=4.8~{\rm GeV}$, $v_b=\sqrt{0.1}$ and $v_c=\sqrt{0.3}$. The decay widths for $T_{4b} \to \pi^+\pi^-$ and $T_{4b} \to K^+K^-$ are listed in TABLE~\ref{gamma-br-table}. The decay widths of $T^{(J)}_{4b}\to \pi^+\pi^-$ and $T^{(J)}_{4b}\to K^+K^-$ are around $10^{-14}$ GeV, which fall far below the current experimental limits.

\section{Summary and Outlook}
\label{section-Summary-Outlook}
In this work, we calculate the decay widths of the fully tetraquark $T_{4c}$ and $T_{4b}$ into light mesons and their antiparticles. The decay width for $T_{4c}$ into pions or kaons is on the order of $10^{-9}~{\rm GeV}$, while for $T_{4b}$, it is approximately $10^{-14}~{\rm GeV}$. 
The branching ratio for $T_{4c}$ decaying into pions or kaons is about $10^{-8}$, which is currently unobservable in experiments.
The exclusive decays of $T_{4c,b}$ into pions or kaons differ from the process $J/\psi \to p\bar{p}$ in that one must retain the $i\varepsilon$-prescription in the denominator due to potential divergences in the integration domain. We have developed a systematic method to tackle these integrals, utilizing the Cheng-Wu theorem and the {\tt QHull} package. This method can be easily extended to other processes, such as $T_{4c,b} \to p\bar{p}$ and various phase-space integrations in high-energy physics.

\begin{acknowledgments}
The authors are very greateful to Long-Bing Chen for his assistance with utilizing the {\tt PolyLogTools}~\cite{Duhr:2019tlz} package.
We also thank Yu Jia and Wen-Long Sang for many useful discussions. The work of F.~F. is supported by the National Natural Science Foundation of China under Grant No. 12275353.
\end{acknowledgments}

\appendix
\section{Triangularization of the Integration Domain}~\label{sec:appendix}

Let us consider the following simple integral ${\cal I}_0$:
\begin{equation}
{\cal I}_0 = \int_{-1}^1 \frac{dx}{x+i\varepsilon} = \int_{-\delta}^\delta \frac{dx}{x+i\varepsilon} = -i\pi.
\label{I0}
\end{equation}
In the sector decomposition method~\cite{Binoth:2003ak,Heinrich:2008si}, to handle the divergence from the denominator, we add $\epsilon$ into the exponents, {\it i.e.},
\begin{equation}
{\cal I}_0 = \lim_{\epsilon \to 0} \int_{-1}^1 dx (x+i\varepsilon)^{-1+\epsilon} = \lim_{\epsilon \to 0} \left[
 \int_{-1}^{-\delta} dx + \int^{1}_{\delta} dx + \int_{-\delta}^{\delta} dx
\right] (x+i\varepsilon)^{-1+\epsilon} .
\end{equation}
Due to the existance of $\epsilon$, the integration from $-\delta$ to $\delta$ becomes zero as $\delta \to 0$. Thus, we have
\begin{equation}
{\cal I}_0 = \lim_{\epsilon \to 0 \atop \delta \to 0} \left[
 \int_{-1}^{-\delta} dx + \int^{1}_{\delta} dx 
\right] (x+i\varepsilon)^{-1+\epsilon}
\equiv \left[
 \int_{-1}^{0} dx + \int^{1}_{0} dx 
\right] (x+i\varepsilon)^{-1+\epsilon} .
\end{equation}
For each integration domain, $x$ has a definite sign, allowing us to explicitly remove the $i\varepsilon$, {\it i.e.},
\begin{equation}
{\cal I}_0 = \int_{-1}^{0} dx (x+i\varepsilon)^{-1+\epsilon} + \int^{1}_{0} dx (x+i\varepsilon)^{-1+\epsilon} 
= (-1+i\varepsilon)^{-1+\epsilon}\int_{-1}^{0} dx (-x)^{-1+\epsilon} + \int^{1}_{0} dx x^{-1+\epsilon} .
\end{equation}
Next, we can integrate over $x$ and obtain:
\begin{equation}
{\cal I}_0 = (-1+i\varepsilon)^{-1+\epsilon} \frac{1}{\epsilon} + \frac{1}{\epsilon} = \frac{1}{\epsilon} \left[ e^{i\pi(-1+\epsilon)} + 1 \right]
= -i\pi + {\cal O}(\epsilon) ,
\end{equation}
showing that we have explicitly recovered the original result in \eqref{I0}. 

The key idea is to sepearte the integration domain into different parts so that the corresponding integrand has a definite sign in each part. 
This approach can be extended to integrals with multiple variables, including those involving $\delta$-functions, with the assistance of the Cheng-Wu theorem~\cite{Cheng:1987ga,Smirnov:2006ry}. 

Now, consider the following general integral, which commonly arises in phase-space integration in high-energy physics:
 \begin{equation}
 	 \int_0^\infty dx_1 \dots \int_0^\infty dx_n \delta(1-\sum_{i=1}^n x_i)\; {\cal P}_1^{n_1} \cdots {\cal P}_m^{n_m},
 \end{equation}
where ${\cal P}_i$ represents a linear combination of $\{x_1,\cdots,x_n\}$. Generally, this type of integral falls under the category of complex-type integrals, which traditional sector decomposition methods with contour deformation are unable to handle.
With the help of the Cheng-Wu theorem~\cite{Cheng:1987ga,Smirnov:2006ry} and {\tt QHull}\footnote{{\tt QHull}: \url{http://www.qhull.org}} package, one can convert this specific type of integration from complex type to Euclidean one, significantly improving the efficiency of numerical integration.
The algorithm is implemented in the {\tt Triangularize} method in the latest version\footnote{{\tt HepLib}: \url{https://github.com/HepLib}} of {\tt HepLib}~\cite{Feng:2021kha}.
 
To illustrate the basic idea, let’s consider the following example:
\begin{equation}
{\cal I}_1 = \int_0^\infty \!\!\!\!dx_1 \int_0^\infty \!\!\!\!dx_2 \int_0^\infty \!\!\!\!dx_3 \; (x_1+x_2-2x_3+i\varepsilon)^{-1+\epsilon} (x_1-x_2-x_3+i\varepsilon)^{-1+\epsilon} \, \delta(1-\sum_{i=1}^3 x_i)
\end{equation}

\begin{figure}[h]
\includegraphics[width=0.35\textwidth]{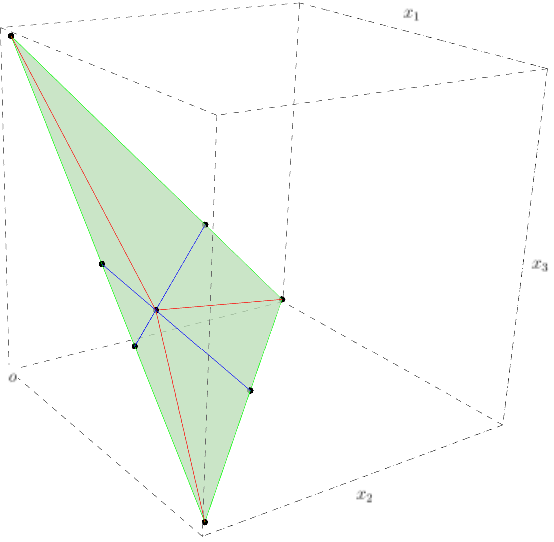}
\caption{Visualization for the triangularization.\label{box}}
\end{figure}

The key point is to divide the integration domain, where $x_i\ge0$, into sub-domains such that the denominators $(x_1+x_2-2x_3)$ and $(x_1-x_2-x_3)$ both have definite signs, $i.e.$, are always positive or negative in each sub-domain. This procedure is analogous to the triangularization of polyhedra in the geometry.
This triangulation becomes more apparent when projecting the entire integration domain $x_{i}\ge0$ onto a specific plane, for example, $x_1+x_2+x_3=1$, represented as the green triangle in FIG.~\ref{box}.

Now the planes $(x_1+x_2-2x_3)=0$ and $(x_1-x_2-x_3)=0$ intersect the green triangle as two line segments, shown as blue lines in FIG.~\ref{box}.
Those two blue line segments divide the triangle into four parts, resulting in both denominators $(x_1+x_2-2x_3)$ and $(x_1-x_2-x_3)$ have definite signs within each division.
To complete the triangulation, we add three additional line segments (shown in red in FIG.~\ref{box}), resulting in the original triangle being divided into seven small triangles. Each small triangle becomes a tetrahedron by connecting to the origin. We can express the three faces of a tetrahedron that pass through the origin in the following general form, and these three faces also correspond to the boundaries of an integral region:
\begin{eqnarray}
\left\{\begin{array}{lll}
a_{11} x_1+a_{12} x_2+a_{13} x_3 = 0 \vspace{0.1cm} \cr
a_{21} x_1+a_{22} x_2+a_{23} x_3 = 0 \vspace{0.1cm} \cr
a_{31} x_1+a_{32} x_2+a_{33} x_3 = 0 
\end{array}\right. 
\end{eqnarray}

We can then perform the following variable change:
\begin{eqnarray}
\left\{\begin{array}{lll}
y_1 = a_{11} x_1+a_{12} x_2+a_{13} x_3 \vspace{0.1cm} \cr
y_2 = a_{21} x_1+a_{22} x_2+a_{23} x_3 \vspace{0.1cm} \cr
y_3 = a_{31} x_1+a_{32} x_2+a_{33} x_3 
\end{array}\right.
\end{eqnarray}

With the help of Cheng-Wu theorem~\cite{Cheng:1987ga,Smirnov:2006ry,Jantzen:2012mw}, which states,
\begin{eqnarray}
{\cal I} = \int_0^\infty d x_1 \cdots \int_0^\infty d x_n \, {\cal F}(x_1,\cdots,x_n) \delta(1-\sum_{i=1}^n a_i x_i),
\end{eqnarray}
where ${\cal F}(\lambda x_1,\cdots,\lambda x_n) = \lambda^{-n} {\cal F}(\lambda x_1,\cdots,\lambda x_n)$, the integral ${\cal I}$ remains unchanged for all cases with nonnegative coefficients and at least one positive coefficient, $i.e.$, ($\forall i, a_i\ge0$ and $\exists k, a_k>0$).
One can transform the integration in each sub-domain into the following form:
\begin{equation}
{\cal I}_1\Big\vert_{\mbox{sub-domain}} = \int_0^\infty \!\!\!\!dy_1 \int_0^\infty \!\!\!\!dy_2 \int_0^\infty \!\!\!\!dy_3 \; ({\cal C}+i\varepsilon)^{-1+\epsilon} ({\cal D}+i\varepsilon)^{-1+\epsilon} \, \delta(1-\sum_{i=1}^3 y_i)
\end{equation}
where ${\cal C}$ and ${\cal D}$ are the two original denominators, expressed as:
\begin{equation}
{\cal C} = c_1 y_1+c_2 y_2+c_3 y_3, \quad {\cal D} = d_1 y_1+d_2 y_2+d_3 y_3 .
\end{equation}
Note that, within a specific sub-domain, both ${\cal C}$ and ${\cal D}$ have definite signs, allowing us to remove the $i\varepsilon$. 
For example, in the case where ${\cal C}>0$ and ${\cal D}<0$, we have:
\begin{equation}
{\cal I}_1\Big\vert_{{\cal C}>0, {\cal D}<0} = e^{i\pi(-1+\epsilon)} \int_0^\infty \!\!\!\!dy_1 \int_0^\infty \!\!\!\!dy_2 \int_0^\infty \!\!\!\!dy_3 \; {\cal C}^{-1+\epsilon} {\cal D}^{-1+\epsilon} \, \delta(1-\sum_{i=1}^3 y_i) .
\end{equation}
Such an integral corresponds to a Euclidean type and can be easily calculated to high precision using the sector decomposition method.
By repeating the procedures described above for all these sub-domains, we arrive at the final numerical integration:
\begin{equation}
{\cal I}_1 = -0.935008093(9) + 0.21230103415(0) i \;,
\end{equation}
which can be compared with the exact analytical result obtained through direct integration using the {\tt Mathematica} program:
\begin{eqnarray}
{\cal I}_1 &=& \frac{{\rm Li}_2\left(-\frac{1}{2}\right)}{6}+\frac{{\rm Li}_2(4)}{6}-\frac{5 \pi^2}{36}+\frac{\ln^22}{12}+\frac{1}{12} \ln3 \ln4+\frac{1}{6} i \pi \ln6 \\
&=& -0.93500809283492521156+0.21230103415326675516 i \;.
\end{eqnarray}
The same idea can be easily extended to integrals involving two $\delta$-functions, such as those in Eqs.~\eqref{T1}, \eqref{T2}, and \eqref{T3} in this work.

\end{document}